\documentclass[epjCONF,onecolumn]{svjour}

\usepackage{graphicx}
\usepackage[varg]{txfonts} 
\usepackage[latin1]{inputenc}

\session-title{Conference Title, to be filled}

\begin{document}

\title{Red Giants from the Pennsylvania - Toru\'n Planet Search}
\author{Pawe\l ~Zieli\'nski\inst{1}\fnmsep\thanks{\email{pawziel@astri.uni.torun.pl}} \and Andrzej Niedzielski\inst{1} \and Monika Adam\'ow\inst{1} \and Aleksander Wolszczan\inst{2,3}}

\institute{Toru\'n Centre for Astronomy, Nicolaus Copernicus University, ul. Gagarina 11, 87-100 Toru\'n, Poland 
\and Department for Astronomy and Astrophysics, Pennsylvania State University, 525 Davey Laboratory, University Park, PA 16802 
\and Center for Exoplanets and Habitable Worlds, Pennsylvania State University, 525 Davey Laboratory, University Park, PA 16802}

\abstract{
The main goal of the Pennsylvania - Toru\'n Planet Search (PTPS) is detection and characterization of planets around evolved stars using the high-accuracy radial velocity (RV) technique. The project is performed with the 9.2 m Hobby-Eberly Telescope. To determine stellar parameters and evolutionary status for targets observed within the survey complete spectral analysis of all objects is required. In this paper we present the atmospheric parameters (effective temperatures, surface gravities, microturbulent velocities and metallicities) of a subsample of Red Giant Clump stars using strictly spectroscopic methods based on analysis of equivalent widths of  Fe~I and  Fe~II lines. It is shown that our spectroscopic approach brings reliable and consistent results.
} 

\maketitle

\section{Introduction}
\label{RG-intro}
Over 350 extrasolar planets are known today and most of them were discovered by the radial velocity (RV) technique. Apart from many surveys searching for exoplanets around solar-like dwarfs, a few surveys looking for planetary companions orbiting evolved stars exist. 
For proper interpretation of the results obtained from RV studies of late-type giants detailed knowledge of their physical parameters is required (e.g. \cite{S06}, \cite{H07}, \cite{T08}). Determination of the atmospheric parameters with high accuracy allows us to place each star on the Hertzsprung-Russell diagram (HRD) and better understand the formation and evolution of the object and its companion.

\begin{figure}
\center
\includegraphics[width=7cm]{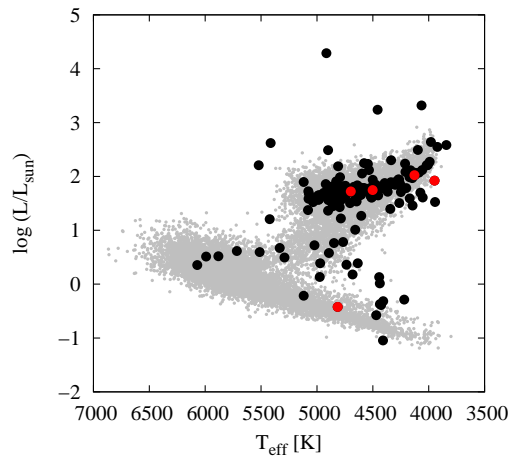}
\caption{HRD for 201 red giants from our sample. Red circles indicate five planetary system hosts from PTPS (\cite{N07}, \cite{N09a}, \cite{N09b}).}
\label{RG-fig1}
\end{figure}

In Fig.~\ref{RG-fig1} we present 201 objects from the Pennsylvania - Toru\'n Planet Search (PTPS), red giants from the Red Giant Clump (RGC) and approximately 11~000 stars from the Hipparcos and Tycho catalogues (gray dots). The effective temperatures and luminosities were taken from \cite{A08}.

\section{Observations and data reduction}
\label{RG-observ}
The observational material used in this work are high-resolution optical spectra of late-type stars observed within the PTPS survey. Most of them are unstudied before objects and only five stars (HD~17092, HD~102272, HD~240210, BD+14~4559 and BD+20~2457) are already known as companion hosting stars (see \cite{N07}, \cite{N09a}, \cite{N09b}).
Observations were obtained with the Hobby-Eberly Telescope (HET, \cite{R98}) equipped with the High Resolution Spectrograph (HRS, R=60~000, \cite{T98}) in the queue scheduled mode (\cite{S07}). The spectrograph was fed with a 2 arcsec fiber. The spectra consisted of 46 Echelle orders recorded on the ''blue'' CCD chip (407.6--592 nm) and 24 orders on the ''red'' one (602--783.8 nm). Typical signal to noise ratio was 200-250 per resolution element. The basic data reduction was performed in a standard manner using IRAF tasks and scripts, which are distributed by the National Optical Astronomy Observatories and operated by the Association of Universities for Research in Astronomy, Inc., under cooperative agreement with the National Science Foundation. 

The equivalent widths (EW) of 283 neutral (Fe~I) and ionized (Fe~II) iron lines from the lists presented in \cite{T05b} (electronic table) were measured with the DAOSPEC software (\cite{P07}) in an automatic manner. In most cases we measured equivalent widths of up to 180 Fe~I and 10 Fe~II lines for further analysis. The estimated accuracy of EW is about 10 \%.

\section{Atmospheric parameters determination}
\label{RG-param}
The atmospheric parameters, i.e. effective temperature  T$_{eff}$, surface gravity log$~\emph{g}$, microturbulence velocity v$_{t}$ and metallicity [Fe/H] for the program stars were obtained with the TGVIT algorithm developed by \cite{T02} and updated in \cite{T05a}. This purely spectroscopic method is based on analysis of iron lines and resulting from three assumptions of LTE that have to be satisfied:\\
--~the abundances derived from Fe~I lines should not show any dependence on the lower excitation potential (excitation equilibrium),\\
--~the averaged abundances from Fe~I and Fe~II lines should be equal (ionization equilibrium),\\
--~the abundances derived from Fe~I lines should not show any dependence on the equivalent widths (matching the curve-of-growth shape).

In our case these three requirements were fulfilled. Lines stronger than 150 m\AA ~at excitation potential of less than 0.5 eV were rejected to avoid a disturbation of the mean iron abundance trend relative to the excitation potential and EWs. Similarly, lines stronger than 200 m\AA ~were discarded in all cases. The final solutions of presented parameters are stable and consistent for all stars included in the sample.

\section{Results}
\label{RG-res}
\begin{figure}
\center
\includegraphics[width=7cm,angle=270]{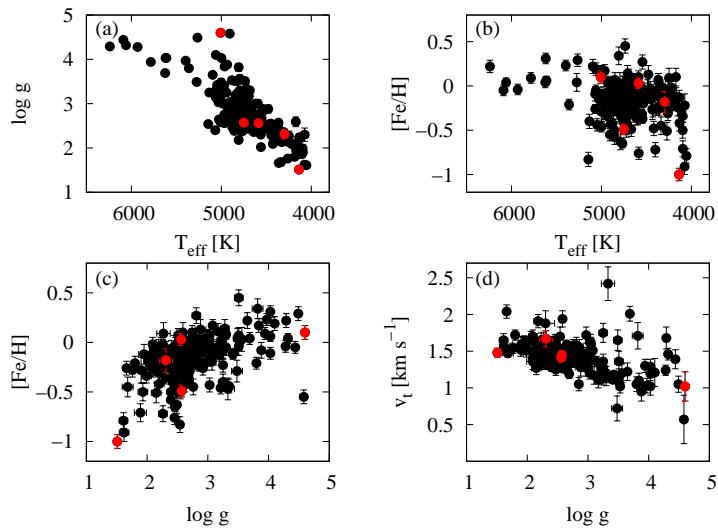}
\caption{Dependencies between four atmospheric parameters for 201 stars from our sample: (a) T$_{eff}$ ~vs.~ log$~\emph{g}$, (b) T$_{eff}$ ~vs.~ [Fe/H], (c) Log$~\emph{g}$ ~vs.~ [Fe/H] and (d) Log$~\emph{g}$ ~vs.~ v$_{t}$.}
\label{RG-fig2}
\end{figure}

The results for the whole sample studied here and relations between four atmospheric parameters are presented on Fig.~\ref{RG-fig2}. Typical average uncertainties of these parameters are $\sigma$T$_{eff} = 15$~K, $\sigma$log$~\emph{g} = 0.06$, $\sigma$v$_{t} = 0.09$~km s$^{-1}$ and $\sigma${[Fe/H]}$= 0.08$. On the figures red dots represent stars with planetary mass companions already detected by our survey. We can conclude that the planet-hosting stars show typical atmospheric parameters for our sample.

\begin{acknowledgement}
\label{RG-acknow}
We thank Y. Takeda, P. Stetson and E. Pancino for making their codes available for us. The Hobby-Eberly Telescope (HET) is a joint project of the University of Texas at Austin, the Pennsylvania State University, Stanford University, Ludwig-Maximilians-Universit\"at M\"unchen, and Georg-August-Universit\"at G\"ottingen. The HET is named in honor of its principal benefactors, William P. Hobby and Robert E. Eberly. 

PZ and MA are supported from the EU Scholarship Programme for PhD Students 2008/2009~--~ZPORR. This work was supported by the Polish Ministry of Science and Higher Education through grant 1P03D 007 30.
\end{acknowledgement}

\end{document}